\documentclass[12pt,a4paper]{article}

\usepackage{amsmath,amssymb}
\usepackage{epsfig}
\usepackage[dvips]{color}

\newcommand{\bu}{$\bullet$\ }
\newcommand\blue{\color{blue}}

\newcommand{\be}{\begin{equation}}
\newcommand{\ee}{\end{equation}}
\newcommand{\bes}{\begin{subequations}}
\newcommand{\ees}{\end{subequations}}
\newcommand{\bea}{\begin{eqnarray}}
\newcommand{\eea}{\end{eqnarray}}
\newcommand{\bear}{\begin{equation}\begin{array}}
\newcommand{\eear}[1]{\end{array}\label{#1}\end{equation}}
\usepackage{wrapfig}

\newlength{\figwidth}
\newlength{\figheight}
\newlength{\twofigheight}
\newlength{\twofigwidth}
\def\gg{$\gamma \gamma$}
 \newcommand{\fn}[1]{\footnote{ #1}}

\begin{document}

\date{\hskip 12cm   IFT 7/2013}

\title{Testing Higgs Physics at the Photon  Collider}

\author{Ilya F. Ginzburg \\
Sobolev Inst. of Mathematics SB RAS and  Novosibirsk University,\\
    630090 Novosibirsk, Russia \\
Maria Krawczyk\\
Faculty of Physics, 
University of Warsaw, Ho\.za 69, 00-681 Warsaw,  Poland}

\maketitle

\begin{abstract}
{{
Here we review  potential of the Photon Collider for  study of Higgs  physics after discovery of the SM-like Higgs boson at the  LHC. In general, the Photon Collider will  fill in the  LHC and ILC results,  giving in some cases  unique  information which cannot be obtained at  other machines.
}}

\end{abstract}

%\today

A Photon  Collider  (hereafter we use abbreviation PLC -- Photon Linear Collider) is based on  photons  obtained  from laser light  back-scattered from high-energy electrons of Linear Collider (LC). Various high energy gamma-gamma and electron-gamma processes can be studied here. With a proper choice of electron beam and laser polarization, the high-energy photons with high degree polarization  (dependent on energy) can be obtained. The direction of this polarization can be easily changed by changing  the direction of {electron and} laser polarization.
By converting  both electron beams to the photon beams one can study $\gamma \gamma$
interactions in the energy range up to $\sqrt{s_{\gamma\gamma}}\sim
0.8 \cdot \sqrt{s_{ee}}$, whereas by converting one beam only the
$e\gamma$ processes can be studied up to $\sqrt{s_{e\gamma}}\sim 0.9 \cdot
\sqrt{s_{ee}}$  \cite{GKST}.

\begin{wrapfigure}[18]{l}{0.7\figwidth}\vspace{-.7cm}
  \includegraphics[width=0.6\figwidth,clip=]{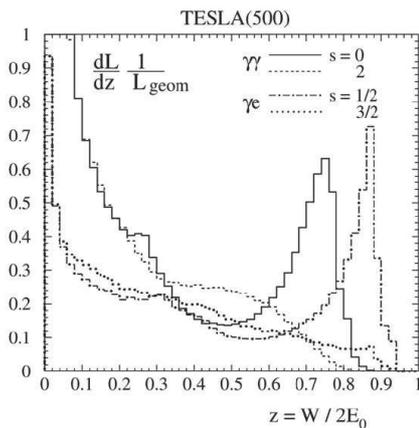}\vspace{-1cm}
  \caption{\it The distribution of $\gamma \gamma$ and $e \gamma $ center-of-mass energy
           $W$ with respect to the $e^+e^-$ energy (2$E_0$) from  simulation of
           the PLC luminosity spectra  \cite{Telnov:1999vz}.  Contributions of various spin states of produced system are shown.}
  \label{fig:spectra}
\end{wrapfigure}
In a nominal LC option, i.e. with the  electron-beam energy of 250~GeV, the geometric luminosity $L_{geom}=12\cdot 10^{34} cm^{-2}s^{-1}$  can be obtained, which is
 about four times higher than  the expected $e^+ e^-$ luminosity.
Still, the luminosity in the high energy $\gamma \gamma$  peak (see
Fig.~\ref{fig:spectra}) corresponds to about $\frac{1}{3}$ of the
nominal  $e^+ e^-$ luminosity -- so we expect $L_{\gamma \gamma}(\sqrt{s_{\gamma\gamma}}> 0.65
\cdot \sqrt{s_{ee}})$ equal to
about 100$fb^{-1}$ per year (400 $fb^{-1}$ for a whole energy
range) \cite{ilc-tdr},\cite{Telnov:1999vz}.
%{\blue In analyses one is typically optimizing the $\gamma \gamma$ energy and polarization  of lasers and electron %beams, while the energy of {\red laser photon is}  assumed to be fixed. }
Adjusting the  initial  electron beam energy and direction of polarizations { of electrons and laser photons at fixed laser photon energy} one can  vary a shape of the \gg \ effective mass spectrum.

At a $\gamma \gamma$ collider the neutral C-even  resonance  with spin 0  can be produced, in contrast to C-odd  spin 1 resonances in the $e^+e^-$ collision. Simple change of signs of polarizations of incident electron and laser photon for one beam transforms PLC to a mode with  total helicity 2 at its high-energy part. It allows to determine degree of possible admixture of state with spin 2 in the  observed  Higgs state.  The  s-channel resonance production of  $J^{PC}=0^{++}$ particle  allows
to perform precise measurement of its properties at PLC.

\bu In summer 2012  a Higgs boson  with mass about  125 GeV  has been discovered at LHC \cite{LHC2012}. We will denote this particle as $\cal H$.  The collected data \cite{ATLAS2013,CMS2013} allow to conclude  that {\it {the SM-like scenario}},  suggested e.g. in \cite{Ginzburg:2001ph,Ginzburg:2003sp},  is realized   \cite{new2011-3}: all measured  $\cal {H}$  couplings are close to their SM-values in their {\it {absolute value}}.  Still     following interpretations of these data are  discussed:  A) ${\cal H}$ is Higgs boson of the SM. B) We deals with phenomenon beyond SM, with
${\cal H}$  being some other scalar  particle (e.g. one of neutral  Higgs bosons of Two Higgs Doublet Model (2HDM) -- in particular MSSM, in the CP conserving 2HDM that are $h$ or $H$). In this approach { following} opportunities are possible:
1)   Measured  couplings  are close to SM-values, however some of them (especially the ttH
 coupling) with  a "wrong" sign 2) In addition some new heavy charged particles, like $H^\pm$ from 2HDM,  can contribute to the loop couplings.
3) The observed signal is not due to {\it one} particle but it is an effect of two or more particles, which were not resolved  experimentally -- {\it the degenerated Higgses.}
 {  Each of these} { opportunities} can lead to  the enhanced or suppressed,    as compared to the SM predictions,   ${\cal H}\gamma\gamma$, ${\cal H}gg$ and ${\cal H}Z\gamma$  loop-coupling.

\bu \ The case with the  observed  Higgs-like signal being due to degenerated  Higgses $h_i$ demands a special effort to diagnose it. In this case the numbers of events with production of some particle $x$ are proportional to sums like $\sum_i (\Gamma^x_i/\Gamma^{tot}_i)\Gamma^{gg}_i$.
Data  say nothing about couplings of the  individual  Higgs particles and   there are no experimental reasons in favor of the SM-like scenario for {\it one} of these scalars.  In such case each of degenerated particles have low total width, and  there is a hope that the forthcoming measurements at PLC can help to distinguish different states due to much better effective mass resolution. The  comparison of different production mechanisms at LHC, $e^+e^-$ LC and PLC will give essential impact in the problem of resolution of these degenerated states. Below we do not discuss the case with degenerated Higgses   with masses $\sim$125 GeV  in more details, concentrating on the case when observed is one Higgs boson $\cal H$, for which the SM-like scenario is realized.

\bu \ In the discussion  we  introduce useful  {\it relative couplings}, defined as ratios of the couplings of each neutral Higgs boson $h^{(i)}$ from the considered model, to the gauge bosons $W$ or $Z$ and to the quarks or leptons ($j=V (W,Z),u,d,\ell...$), to the corresponding SM couplings:
$ \chi_j^{(i)}=g_j^{(i)}/g_j^{\rm SM}$.  Note that all couplings to EW gauge bosons $\chi_V^{(i)}$ are real, while the couplings to fermions are generally complex. For CP-conserving case of 2HDM we have in particular $\chi_j^h$, $\chi_j^H$, $\chi_j^A$ (with $\chi_V^A=0$), where couplings of fermions to $h$ and $H$ are real while couplings to $A$ are purely imaginary.

{\it The SM-like scenario}  for the observed  Higgs  ${\cal H}$, to be identified with some neutral $h^{(i)}$,  corresponds to $|\chi_j^{\cal H}|\approx 1$.  Below we assume this scenario is realized at present.

\bu It is known already from a long time that the PLC is  very good observatory of the scalar sector of the SM and beyond SM, leading to important  { and in many cases} complementary to the $e^+e^-$ LC case tests of the EW symmetry breaking mechanism \cite{GinHiggs}-\cite{Tel2012}. The $e^+e^-$ LC, together with its  PLC  options ($\gamma \gamma$ and $e \gamma$),  is very well suited for the precise study of properties of this newly discovered ${\cal H}$ particle, and other scalars. In particular, the
PLC offers a unique opportunity to study resonant production of Higgs bosons  in the process $\gamma \gamma \rightarrow {\rm  Higgs}$ which is sensitive to  charged fundamental particles of the theory.  In principle, PLC
allows  to study also resonant production of heavier neutral Higgs particles from the extension of the SM. Other physics topic which could be studied  well at PLC is the CP property of Higgs bosons. Below
we discuss the most important aspects of the Higgs physics  which can be investigated at PLC. Our discussion is based on analyses  done during last two decades  and takes into account also some recent "realistic" simulations supporting those results.\\

\centerline{\large\bf  I. Studies of 125 GeV Higgs $\pmb{\cal H}$}
\vskip 0.5cm
 The  discussion in this section is related to the case when $\cal H$ is one of the Higgs bosons $h^{(i)}$ of 2HDM.  In the CP conserving case of 2HDM it can be either $h$ or $H$.

\begin{wrapfigure}[17]{r}{0.45\textwidth}\vspace{-1cm}
\includegraphics[height=\twofigheight]{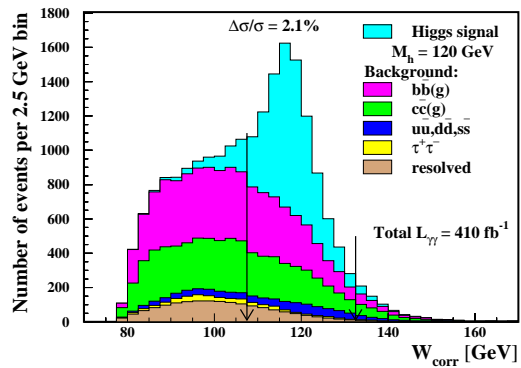}\vspace{-0.4cm}
 % \end{center}
 \caption{\it
Distributions of the corrected invariant mass, $W_{corr}$, for
selected $b \bar{b}$ events; contributions of the signal, for $M_{H_{SM}} =
$ 120~GeV, and of the different  background processes, are shown separately \cite{Niezurawski:2005cp}.}
 \label{fig:phys_h1}
 \end{wrapfigure}

%\underbrace{}
$\bullet$ \   Several NLO analyses of the production at the PLC of a light SM-Higgs boson $H_{SM}$ decaying into $b \bar b$ final state were performed, including the detector simulation, eg.~\cite{JikS}--\cite{SM-Mayda}. These analyses demonstrate a high  potential of this collider to measure accurately the Higgs two-photon width. By combining the production rate for $\gamma \gamma \rightarrow H_{SM}\rightarrow b \bar b$ (Fig. \ref{fig:phys_h1}), to be measured with 2{\blue.1} \% accuracy,  with the measurement of the $Br(H_{SM}\rightarrow bb)$ at
$e^+e^-$ LC, with accuracy ~$\sim$ 1~\%, the width $\Gamma(H_{SM} \rightarrow\gamma \gamma)$ for $H_{SM}$ mass of 120 GeV can be determined with precision $\sim$ 2 \%. This can be compared to the present value  of the measured at LHC signal strength for 125 GeV $\cal H$ particle,  which ratio to the expected signal for SM Higgs with the same mass  (approximately equal to the ratio of $|g_{\gamma \gamma {\cal H }}|^2/|g_{\gamma \gamma {H_{SM} }}|^2$),    are 1.55$+$0.33$-$0.28 and 0.78 $\pm0.28/0.26$ from ATLAS \cite{ATLAS2013} and CMS \cite{CMS2013}, respectively.

\bu \ The process $\gamma\gamma\to {\cal H}\to \gamma\gamma$ is also observable at the PLC with reasonable rate \cite{SM-Mayda}. This measurement allows to measure directly two-photon width of Higgs without assumptions about unobserved channels, couplings, etc.

\bu \  Neutral Higgs resonance  couples to photons via loops with
charged particles. In the  Higgs $\gamma \gamma$ coupling  the heavy
charged particles, with masses generated by the Brout-Englert-Higgs-Kibble mechanism, do not decouple. Therefore the ${\cal H}\to\gamma \gamma$ partial width  is sensitive to the contributions of charged particles with masses even far beyond the energy of the $\gamma \gamma$ collision.  This allows to recognize  which type of extension of the  minimal SM is realized.  The $H^+$ contribution to the ${\cal H} \gamma \gamma$ loop coupling   is proportional to  ${\cal H} H^+H^-$ coupling,   which value and sign
can be treated as
free parameters of model\footnote{Except if some additional symmetry is present in the model.}.%\hspace{-0.5cm}
The simplest  example gives a 2HDM with Model II Yukawa interaction (2HDM II). For a small $m_{12}^2$ parameter
the contribution of the charged Higgs boson $H^+$ with mass  larger than 400 GeV  leads to 10\% suppression in the ${\cal H}\to\gamma \gamma$ decay width as compare to the SM one, for $M_{\cal H}$ around 120 GeV \cite{Ginzburg:2003sp,Ginzburg:2001ph}, Table \ref{table1} (solution A).
%Therefore
The enhancement or decreasing of  the ${\cal H}  \gamma \gamma$ coupling is possible, as discussed for 2HDM with various Yukawa interaction models in \cite{Bernal:2009rk}--\cite{posch} as well in the  Inert Doublet Model\fn{ That is
the $Z_2$ symmetric 2HDM where one  Higgs doublet  plays a role of SM Higgs field $\phi_S$, interacting with fermions as in Model I, with  the  SM-like Higgs boson $h$ and another Higgs doublet  $\phi_D$, having no v.e.v.. The latter one  contains  four  scalars $D, \,D^A,\,D^\pm $, the lightest among them $D$ (analog of  $H$ of 2HDM) can be DM particle,  scalars $D^A$ and   $D^\pm$ (analog of $A$ and $H^\pm$, respectively).} \cite{arh,bs}.

\begin{wrapfigure}[15]{r}{0.46\textwidth}\vspace{-0.55cm}\hspace{-0.5cm}
\includegraphics[height=0.3\textheight,width=0.35\textwidth,angle=270]{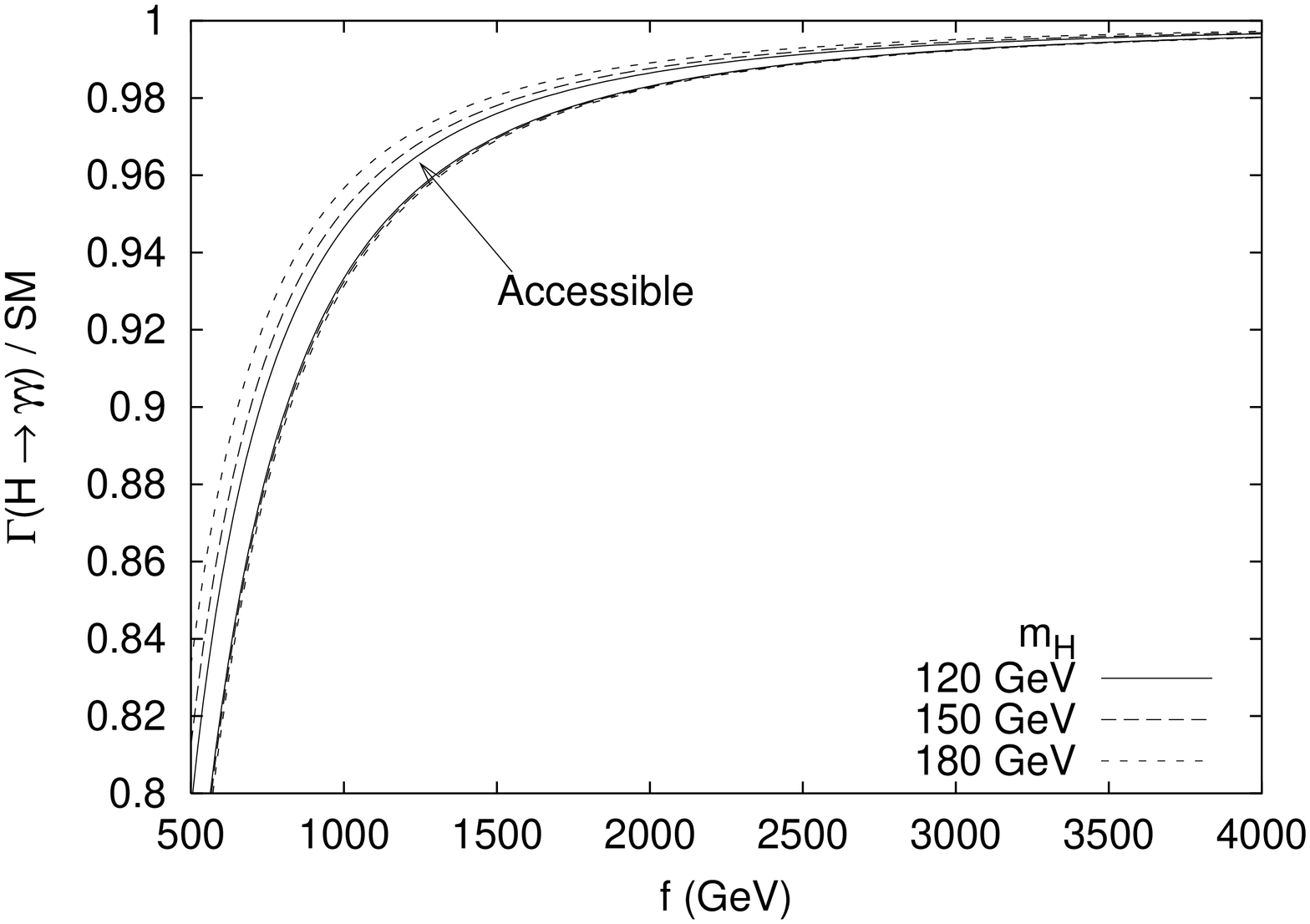}\vspace{-0.4cm}
 % \end{center}
 \caption{\it
Ratio   $\dfrac{\Gamma(h \rightarrow \gamma \gamma)}{\Gamma(h\to \gamma\gamma)^{SM}}$
as a function of the mass scale of the new physics
$f$ in the Littlest  Higgs model \cite{Han:2003gf}, for different Higgs boson masses.
}
 \label{fig:little}
 \end{wrapfigure}

In the Littlest Higgs model a 10\% suppression of the $\gamma \gamma$ decay width for $M_{\cal H}\approx 120$ GeV is expected due to the  new heavy particles with mass around 1 TeV  at the suitable scale of couplings for these new particles \cite{Han:2003gf}, \cite{Huang:2009nv},
 Fig.~\ref{fig:little}.
\vskip 2cm

\bu \ The  Higgs $\gamma \gamma$ loop coupling is sensitive to the relative signs of various contributions. For example,  in 2HDM II  sign of some  Yukawa couplings  may differ from the SM case, still strength (ie. absolute value) of all squared direct Higgs couplings to WW/ZZ and fermions  being as in the SM. This  may lead to the enhancement of the ${\cal H} \to \gamma \gamma$ decay-width with respect to the SM predictions, up to 2.28 for a "wrong" sign of the ${\cal H} tt$ for $M_{\cal H}=120$  GeV  (1.28 for  ${\cal H} \to gg$ and 1.21 for ${\cal H}  \to Z\gamma$, respectively)  coupling, Table \ref{table1}  (solution $B_{{\cal H} t}$), \cite{Ginzburg:2001ph}\fn{The recent analysis of the LHC data     leads to constraints of the relative ${\cal{H}}tt$ coupling $\chi_t^{\cal{H}}$
\cite{mele}.}.
The "wrong" sign of ${\cal H}bb$ coupling (solution $B_{{\cal H} b}$ in Table \ref{table1}) could lead to a enhancement in the ${\cal H} \to gg$,  and in the corresponding rate for gluon fusion of Higgs at LHC, similarly as  the "wrong" sign of ${\cal H}tt$ coupling.  Such solution is still considered as a possible for 125 GeV ${\cal H}$ particle \cite{ATLAS2013}.

\begin{table}[ht]
\begin{center}
\begin{tabular}{||c|c|c|c|c|c||}
\noalign{\vspace{-9pt}}
\hline\hline
solution&basic couplings&$|\chi_{gg}|^2$ &
$|\chi_{\gamma\gamma}|^2$ & $|\chi_{Z\gamma}|^2$\\ \hline\hline
$A_{{\cal H}}$
&$\chi_V\approx\chi_b\approx\chi_t\approx \pm1$
& 1.00 & 0.90 & 0.96 \\ \hline
$B_{{\cal H} b}$
&$\chi_V\approx-\chi_b\approx\chi_t\approx \pm1$
& 1.28 & 0.87 & 0.96 \\ \hline
$B_{{\cal H} t}$
& $\chi_V\approx\chi_b\approx-\chi_t\approx \pm1$
& 1.28 & 2.28 & 1.21 \\ \hline\hline
\end{tabular}
\end{center}
\vspace*{-3mm}
\caption{\it SM-like realizations in the 2HDM~II  \cite{Ginzburg:2001ph},\cite{Ginzburg:2003sp} together
with ratios of loop-induced partial widths to their SM values at
$M_{\cal H}=120$~GeV, $M_{H^{\pm}}=
 $800~GeV, $|m_{12}^2|\le 40$~GeV$^2$. }
\vspace*{-2mm}
\label{table1}
\end{table}

\bu   The observed Higgs particle can have definite CP parity or can be admixture of states with different CP parity  ({\it  {CP-mixing}}). In the latter case the PLC provides the best among  all colliders place for the study of such mixing. Here, the  opportunity to simply vary polarization of photon beam allows to study this mixing via dependence of the production cross section on the incident photon polarization \cite{Grzadkowski:1992sa},\cite{HCPH},\cite{gii},\cite{Zerwas-CP}. In particular, the change of sign of circular polarization ($++ \leftrightarrow --$) results in variation of production cross section of the 125 GeV Higgs in 2HDM  by up to about 10\%, depending on a degree of CP-admixture. Using mixed circular and linear polarizations of photons gives opportunity  for more detailed investigations.

\bu \   The important issue   is to measure  a  Higgs selfcoupling, ${\cal H}{\cal H}{\cal H}$. In the SM this selfcoupling is precisely fixed via Higgs mass (and v.e.v. $v=246$~GeV), while deviations from it's SM value would be  a clear  signal of more complex Higgs sector. Both at the $e^+e^-$ collider
and at the \gg \ collider the  two neutral Higgs bosons are produced  in processes  both with and without  selfinteraction, namely
$$\begin{array}{c}
e^+e^-\to Z\to {{\cal H}(Z\to Z{\cal H})} \oplus e^+e^-\to Z\to {Z({\cal H}\to {\cal H}{\cal H})};\\[2mm]
\gamma\gamma\to loop\to { {\cal H}{\cal H}}\oplus \gamma\gamma\to loop\to  {{\cal H}\to {\cal H}{\cal H}}.\end{array}
$$
In the SM case  the cross sections for above processes are rather low  but measurable, so that coupling under interest can be extracted,  both in the $e^+e^-$ and \gg\,\, modes of $e^+e^-$ LC, see \cite{Belusevic:2004pz}-\cite{Tsumura:2011zz}.
The feasibility of this measurement  at a PLC has been performed recently in \cite{Kawada:2012uy}. For Higgs mass of 120 GeV and the integrating luminosity 1000 fb$^{-1}$ the  statistical sensitivity as a function of the $\gamma \gamma$ energy for measuring the deviation from the SM Higgs selfcoupling $\lambda=\lambda_{SM} (1+\delta \kappa)$ has been estimated. The optimum  $\gamma \gamma$ collision  energy  was found  to be around 270 GeV for a such Higgs mass,   assuming that  large  backgrounds due to $WW/ZZ$ and $bbbb$ production  can  be suppressed for correct  assignment of tracks.   As a result, the  Higgs  pair  production  can  be  observed  with  a statistical significance  of  5 $\sigma$ by operating the PLC  for 5 years.
%These estimates demonstrate  low,  sensitivity of these measurements to the  deviations from the  SM.

\bu \ The smaller but interesting effects are expected in $e\gamma\to e{\cal H}$ process with\linebreak[4] $p_{\bot e}> 30$~GeV, where ${\cal H}Z\gamma$ vertex can be extracted with reasonable accuracy \cite{Gvych}.\\

\centerline{\large\bf II. Studies of  heavier Higgses, for  125 GeV $\pmb{{\cal H}=h^{(1)}}$}

\vskip 0.5cm
A direct discovery of other Higgs bosons and measurement of their couplings to gauge bosons and fermions is necessary for clarification the { way the SSB is realized}. In this section we consider the case when observed 125 GeV Higgs is the lightest neutral Higgs, ${\cal H}=h^{(1)}$ (in particular in the CP-conserving case this means ${\cal H}=h$).  A single Higgs production at \gg \ \  collider allows to  explore roughly the same  mass region for neutral Higgs bosons at the parent $e^+e^-$ LC but with higher cross section and lower background. The $e\gamma$ collider allows in principle to test wider mass region in the process $e\gamma\to eH, eA$  however  with a  lower cross section.

\bu {Before general discussion, we  present some properties of one of the simplest Higgs {} { model} beyond the minimal SM, namely  2HDM (in particular, also the Higgs sector of MSSM), having in mind that the modern data are in favour of a SM-like scenario.
Let us enumerate here some important properties of 2HDM for each neutral Higgs scalar $h^{(i)}$ in the CP conserving case $h^{(1)}=h$, $h^{(2)}=H$, $h^{(3)}=A$:
\begin{itemize}
\item[{\bf A.}]  {\it For an arbitrary Yukawa interaction} there are sum rules for coupling of different neutral Higgses to gauge bosons $V=W,\,Z$ and to each separate fermion $f$ (quark or lepton)
\be
\sum\limits_{i=1}^{3} (\chi_V^{(i)})^2=1\,.\qquad\sum\limits_{i=1}^{3}(\chi_f    ^{(i)})^2=1\,.
\label{srW}
\ee
The first  sum rule (to the gauge bosons) was discussed  e.g. in \cite{gunion-haber-wudka}--\cite{GK05}. The second one was obtained only for Models I and II of Yukawa interaction \cite{Grzadkowski:1999wj}, however in fact it holds for any Yukawa sector \cite{GKr2013}.

In the first sum rule all quantities  $\chi_V^{(i)}$ are real. Therefore,  in SM-like case  (i.e. at $|\chi_V^{(1)}|\approx 1$) both    couplings $|\chi_V^{2,3}|$  are small. The couplings entering the second sum rule (for fermions) are generally complex. Therefore this sum rule  shows that for $|\chi_f^{(1)}|$ close to 1, {either  $ \left|\chi_f^{(2)}\right|^2$ and $\left|\chi_f^{(3)}\right|^2$ are simultaneously small, or  $ \left|\chi_f^{(2)}\right|^2 \approx\left|\chi_f^{(3)}\right|^2$}.

\item
[{\bf B.}]  For the 2HDM I   there are  simple relations, which  in the CP conserved case  are
as follows
\be
\chi_u^{(h)}=\chi_d^{(h)}\,,\qquad \chi_u^{(H)}=\chi_d^{(H)}\,.
\label{VodIeq}
\ee
\item
[{\bf C.}]  In the 2HDM II following relations hold:\\
a) {\em The pattern relation} among the relative couplings
for  {\it each neutral Higgs particle $h^{(i)}$}
\cite{Ginzburg:2001ss,Ginzburg:2002wt}:
 \bes\label{ModIIeq}\be\label{2hdmrel}
(\chi_u^{(i)} +\chi_d^{(i)})\chi_V^{(i)}=1+\chi_u^{(i)}
\chi_d^{(i)}\,.
 \end{equation}
b) For each neutral Higgs boson $h^{(i)}$ one can write a horizontal  sum rule
\cite{Grzadkowski:1999ye}:
\begin{equation}
|\chi_u^{(i)}|^2\sin^2\beta+|\chi_d^{(i)}|^2\cos^2\beta=1\,.\label{srules}
\end{equation}\ees
\end{itemize}

\bu   Below, in Table 2, we present benchmark points for
the SM-like $h$ scenario in the CP conserving 2HDM II. The total widths for H and A  for various $\chi_t^A=1/\tan\beta$ are shown assuming
with  $\chi_V^h\approx 0.87 $, $|\chi_V^H|=0.5$ and  $|\chi_t^h|=1$ for $H$ and $A$ .   \footnote{The total width  $\Gamma_H$ differs from
the total width  $\Gamma_A$ by the $W/Z$ contribution,  since $\chi^A_V=0$.}
\begin{table}[htb]
\begin{center}
\begin{tabular}{|c|c||c|c|c|}\hline
&$\Gamma_H, \qquad \Gamma_A$&$\Gamma_H, \;\; \Gamma_A$&$\Gamma_H, \;\;
\Gamma_A$\\\hline
$M_{H,A}$& $\tan\beta=1/7$ &$\tan\beta=1$ &
$\tan\beta=7$ \\\hline
200& $0.35\qquad 8\cdot 10^{-5}$&$0.35\qquad 4\cdot 10^{-3}$&$0.4\qquad 0.2$\\\hline
300 &$2.1\qquad 1.2\cdot 10^{-4}$&$2.1\qquad 6\cdot 10^{-3}$&$0.75\qquad 0.3$\\\hline
400& $138\qquad 132$&$8.8\qquad 2.7$&$2.5\qquad
0.45$\\\hline
500& $537\qquad 524$&$22.8\qquad 10.7$&$6.1\qquad
0.7$\\\hline
\end{tabular}
\end{center}
\label{benchtab}
\caption{\it Total width (in MeV) of $H$, $A$ in some benchmark points for the SM-like $h$ scenario ($M_h$=125 GeV)  in the 2HDM  ($\chi_V^h\approx 0.87 $, $|\chi_V^H|=0.5$ and  $|\chi_t^h|=1$).
Results for  $\tan \beta=1/7, \, 1 \ \  and \ \  7$ are shown.
}
\end{table}

%One can see that in
In the SM-like $h$  scenario it is follows from sum rule \eqref{srW} that the $W$-contribution to the $H\gamma \gamma$
 width is much smaller than that  of would-be heavy SM Higgs, with the same mass, $M_{H_{SM}}\approx M_H$. At the
large $\tan\beta$ also $H\to tt$, $A\to tt$ decay widths  are extremely
small, so that the total widths of $H$, $A$ become very small \fn{At  $\tan\beta\ll 1$ we obtain the  strong interaction in the Higgs sector mediated by $t$-quarks, what is signalizing by the  fact that the  calculated in standard approach total widths of heavy $H$, $A$ is becoming  close to or even higher than the corresponding masses.
Of course,  in this case such  tree-level estimates become inadequate. In the same manner at $\tan\beta>70$ corresponds to the region of a strong interaction in the Higgs sector mediated by $b$-quarks. We don't consider such scenarios.}.

\bu \ Let us compare properties of heavy $H$, $A$  in 2HDM  with a  would-be heavy SM Higgs-boson with the same mass.
The cross section for production of such particles  in the main gluon-gluon fusion channel,  being $\propto \Gamma_{H,A}^{gg}\Gamma_{H,A} /M_H^3$,   is lower than that in SM. At large $\tan\beta$ resonances $H,A$ become very narrow, as discussed above,  besides, the two-gluon decay width become  about $1/\tan^2\beta$ smaller. So, in this main at LHC  production channels cross section are roughly $1/\tan^4\beta$ smaller than that for the would-be SM Higgs boson with the same mass and   $H$ and $A$ can escape observation in these channels at LHC. (The same is valid for $e^+e^-$ LC due to small value of $\chi_V^H$ for $H$ and $\chi_V^A=0$.)

\begin{figure}[htb]
 \includegraphics[height=.4\textheight,width=0.5\textwidth]{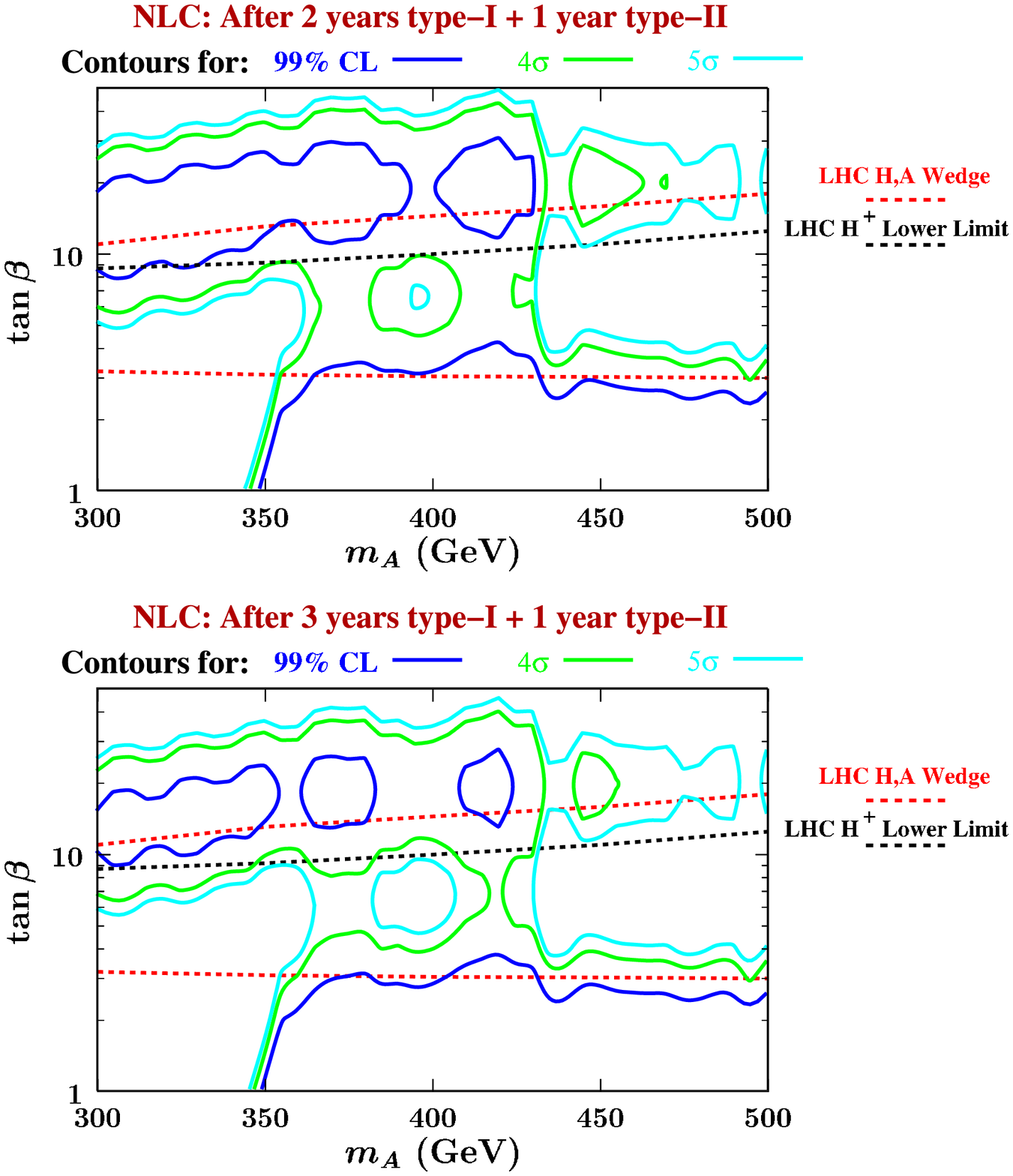}\hspace{7mm}
  \includegraphics[width=0.5\textwidth,height=0.3\textheight]{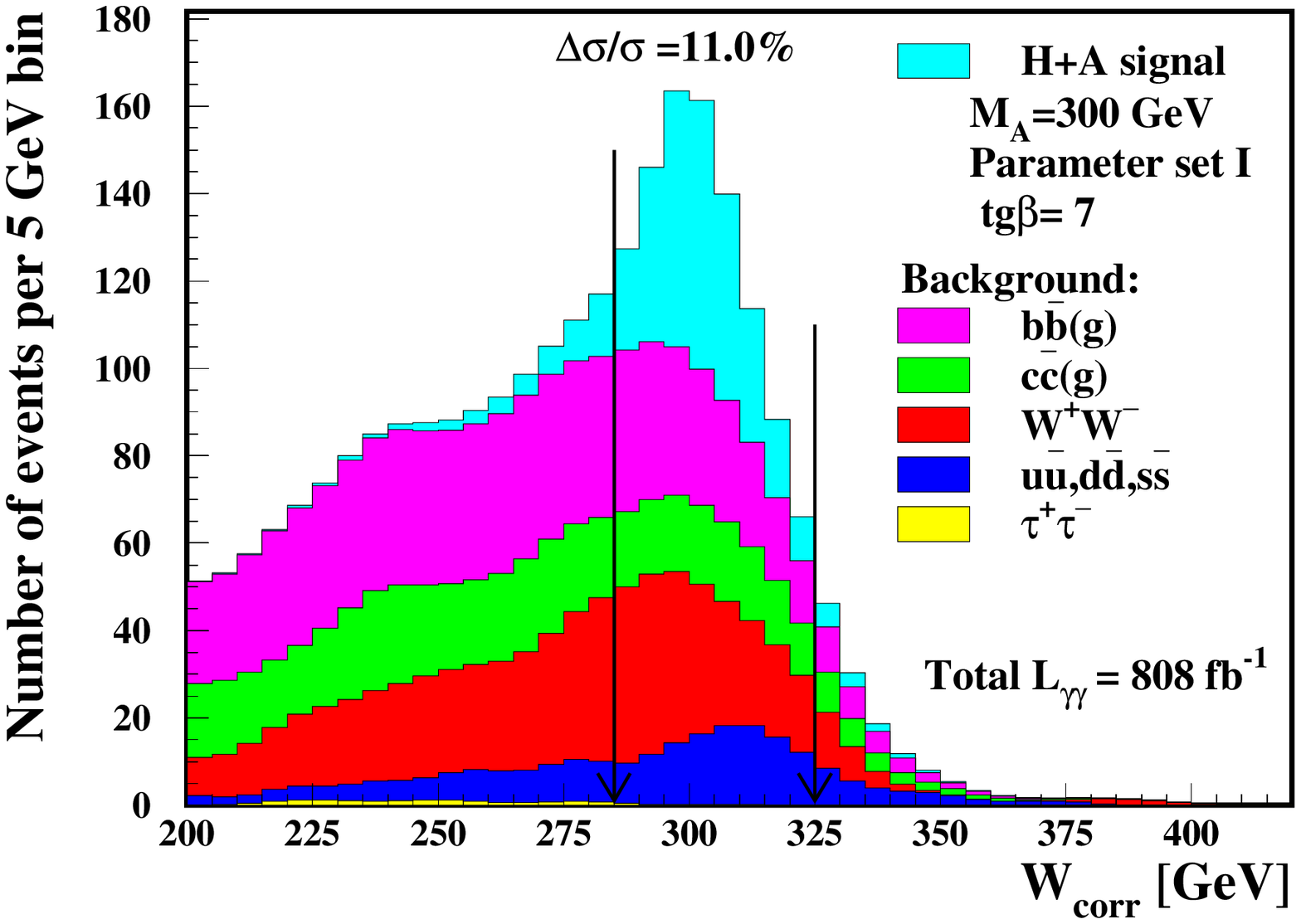}
  \caption{\it
Left: Production of  A and H, with  parameters  corresponding to
 the LHC wedge, at the $\gamma \gamma$ collider.
 Exclusion and discovery limits obtained for NLC collider for
$\sqrt {ee} =$630 GeV, after 2 or 3 years of operation \cite{Asner:2001ia},
 Right: The case $M_H=M_A=300$ GeV at $\chi_V^H\approx 0$ in the MSSM. Distributions of the corrected invariant mass $W_{corr}$ for selected $b \bar{b}$ events  at   $\tan \beta=7$ \cite{nzk2005}.}
%Right: Statistical significance of the Higgs-boson %production measurement as a function of $\tan\beta$ at %different MSSM parameters %$\mu,\,M_2,\,A_{\tilde{f}},\,M_{\tilde{f}}$ \cite{}
   \label{fig:wedge}
\end{figure}

Moreover, in MSSM  with $M_h=125$~GeV we { can } have heavy and degenerate $H$ and $A$, $M_H\approx M_A$. At large $\tan\beta$  discvery channel  of $H/A$ at LHC is $gg\to b\bar{b}\to b\bar{b}H/A$. Nevertheless,  in some region of parameters, at intermediate $\tan\beta$},  { these $H{\, {\rm and}\,}\,A$ are elusive at LHC.  That is so called {\it LHC wedge region}  \cite{wedgeHein}, see the  latest analysis  \cite{Carena:2013qia}.
The PLC allows to diminish this region of elusiveness, since  here the $H$ and $A$ production is generally not { strongly} suppressed and the $b\bar {b}$ background is under control \cite{zerwas0,Asner:2001ia,nzk2005,Spira:2006aa}.  The figs.~\ref{fig:wedge} show that PLC allows to observe joined effect of $H,\,A$ within this wedge region. Precision
between 11\% to 21 \% for $M_A$ equal to 200-300 GeV,  $\tan\beta$ = 7 of the Higgs-boson production measurement ($\mu$ =200 GeV and $A_f$=1500 GeV) can be reached after one year \cite{nzk2005}.  To separate these resonances even in the limiting case $\chi_V^H=0$ is a difficult task,
%Note that in the more realistic case $\chi_V^H\neq 0$ (e.g. 0.3-0.4) the production cross section of $H$ at PLC increases, and signal, shown in this figure increases. In this case the observed will be practically $H$, and $A$ will be hardly observable in its shadow.
since the total number of expected events  is small.
%, {\bf however not hopeless, as we  discuss below}.
% so that polarization studies at PLC {\bf  should} help  to separate  $H$ and $A$ signals, see below.

\bu  At $\chi_V^H\neq 0$, equal say  0.3-0.4  (what is  allowed by current  LHC measurement of  couplings of ${\cal H} = h$ to $ZZ$),   an  observation of $H\to ZZ$ decay channel can be good method for the  $H$ discovery in  2HDM.
The signal $\gamma\gamma\to H\to WW, ZZ$ interferes with background of $\gamma\gamma\to WW, ZZ$, what results in irregular structure in the effective-mass distribution of products of reaction $\gamma\gamma\to WW, ZZ$ (this interference is constructive and destructive below and above resonance, respectively). The study of this irregularity seems to be the   best method for discovery of heavy Higgs, decaying  to $WW,\, ZZ$  \cite{GIvWWH},  and to  measure the  corresponding $\phi_{\gamma\gamma}$ phase, provided it couples to ZZ/WW reasonably strong\fn{Similar calculations given in  \cite{Niezurawski:2002jx} demonstrate this opportunity for a  2HDM version $B_{hu}$. %(Fig. 3).
}.

\begin{figure}
\begin{center}
\hspace*{3cm}
\includegraphics[width=0.33\textwidth,height=0.18\textheight]{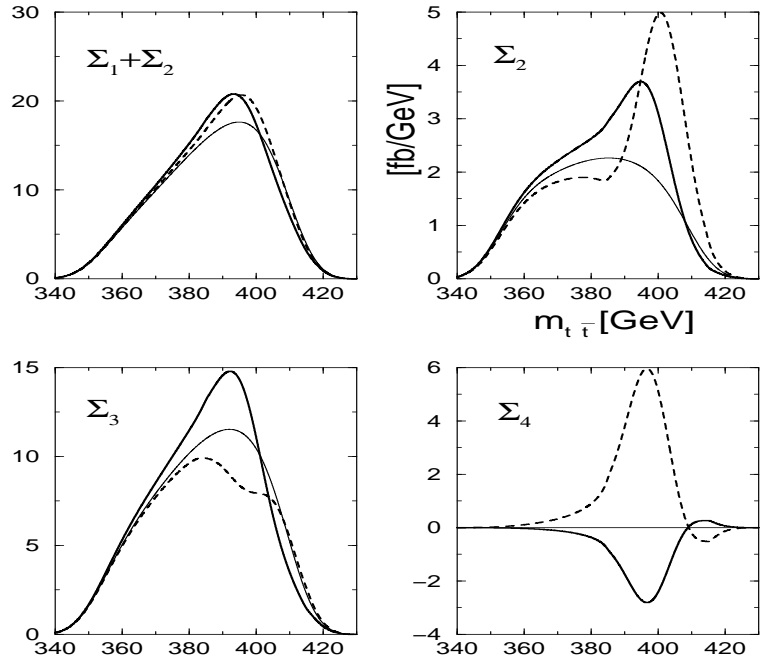}\vspace{4cm}
 \caption{\it The specific  decay angular distributions $\Sigma_i$ in the $\gamma \gamma \rightarrow h^{(i)}\to t\bar{t}$ process in dependence on the $t\bar{t}$ invariant mass for the scalar (dashed)   and pseudoscalar (thick solid) $h^{(i)}$ with $M_H= 400$ GeV   \cite{Asakawa:2003dh}.}
 \label{fig:phys_h2}
\end{center}
 \end{figure}

\bu \   Just as it was described above for the observed 125 GeV  Higgs, PLC provides the best among  colliders place for the study of spin and the CP properties of heavy $h^{(2)}$, $h^{(3)}$. That are CP parity in the CP conserved case  (with ($h^{(2)}$, $h^{(3)}$ = ($H,\,A$)), and (complex) degree of   the admixtures of states with different CP parity, if CP is violated. This admixture determines dependence  on the Higgs production cross section  on direction of incident photon polarization \cite{HCPH}--\cite{Zerwas-CP}, \cite{Choi:2002jk,Ellis:2004hw}. These polarization measurements are useful in the study of the case when the heavy states $h^{(2)}$, $h^{(3)}$ ($H,\,A$) are degenerated in their masses.
{ {A study \cite{nzk0710} shows that the 3-years operation of PLC with linear polarization of photons,  the production cross-section of the $H$ and $A$ corresponding to the LHC wedge for MSSM (with mass ~300 GeV) can be separately measured with precison 20\%.
Pure scalar versus  pure pseudoscalar states can be distinguished at 4.5 $\sigma$ level}}.

We point out on important difference between the $CP$ mixed and the mass-degenerate states.  In the degeneracy of some resonances $A$ and $B$ one should distinguish two opportunities:
\begin{itemize}
\item[a)] instrumental degeneracy when  $|M_B-M_A|>\Gamma_B+\Gamma_A$, with mass difference  within a mass resolution of detector. This effect can be resolved with improving of a resolution of the detector
\item[b)] physical degeneracy when  $|M_B-M_A|<\Gamma_B+\Gamma_A$.
\end{itemize}
In the CP conserving case for both types of degeneracy the overlapping of $H,\,A$ resonances does not result in their mixing, and
the production of a resonante   state cannot vary with change  of sign of photon beam polarization. In the CP violating  case,  the overlapping of resonances results in additional mixing of incident $h^{(2)}$, $h^{(3)}$ states, and the production cross-section varies with  the change of  polarization direction of incident photons.

\bu \ Another method for study of CP content of a produced particle provides the measurement of angular distribution of  decay products.
%in the decay of this  $H$.
In the  $t \bar t$ decay mode one can perform a study of the CP-violation, exploiting fermion polarization.
The interference between the Higgs exchange and the continuum amplitudes can be sizable for the polarized photon beams, if helicities of the top and anti top quarks are measured. This enables  to determine the CP property of the Higgs boson completely \cite{Asakawa:2000jy,Asakawa:2003dh},  Fig. \ref{fig:phys_h2}.

\bu \   The discovery of charged Higgses $H^\pm$ will be a crucial signal of the BSM form of Higgs sector. These particles can be produced both at  the $e^+e^-$ LC ($e^+e^-\to H^+H^-$) and at  the  PLC ($\gamma\gamma \to H^+H^-$). These processes  are described well by QED. The $H^+H^-$ production process at PLC has worse energy-threshold behaviour than the corresponding process at the $e^+e^-$ LC, but higher cross section. On the other hand, the process $e^+e^-\to H^+H^-$ can be analysed at LC better by measurements  of decay products due to known  kinematics. At the PLC the variation of a initial-beam polarization could be  used for checking up spin of $H^\pm$   \cite{ilya2}. See also analysis for Model III  in  \cite{Martinez:2008hu}.

 \bu \  After a  $H^\pm$ discovery,  the observation of processes $e^+e^-\to H^+H^-h$ and $\gamma \gamma \to H^+H^-h$, $H^+H^-H$, $H^+H^-A$  may provide direct  information on  a triple Higgs ($H^+H^-h$) coupling $\lambda$,  with  cross sections in both cases  $\propto \alpha^2\lambda^2$.
The \gg \,\, collisions are preferable here due to a substantially higher cross section and opportunity of study   polarization effects in the production process via variation of initial photon polarizations.

\bu \  Synergy of  LHC, $e^+e^-$ LC and PLC  colliders may be useful in determination of  Higgs couplings, as different production processes dominating at these colliders  lead to different sensitivity to gauge and Yukawa couplings. For example  $e^+e^-$ LC Higgstrahlung leads to  large sensitivity to the Higgs coupling to the EW gauge bosons, while at  PLC $\gamma \gamma$ and $Z\gamma$  loop couplings depend both on the Higgs gauge and Yukawa couplings, as well as on coupling with $H^+$,  see results both for  CP conservig/CP violating in e.g. \cite{Godbole:2006eb,Heinemeyer:2005uf, Niezurawski:2006hy}.
\vskip 1cm

We thank  Gudi Morgat-Pick and Peter Zerwas  for a suggestion to prepare this overview for ILC Forum 2012. We are grateful to Filip \.Zarnecki for clarification of old analyses as well as to  Jan Kalinowski.
The work was partly supported by  Polish National Center for Science, grant 
NCN OPUS 2012/05/B/ST2/03306 (2012- 2016), and  by  BMBF, DAAD PPP Poland Project 56269947, "Dark Matter at Colliders"  (M. K.),  grants RFBR 11-02-00242,
NSh-3802.2012.2 (I.G.).

%\vspace{-0.5em}

\end{document}